%% file: M104_paper_with_referee_edits.tex
\documentclass[10pt,preprint2,longabstract]{aastex}

\shorttitle{GC Kinematics of M104}
\shortauthors{Dowell et al.} 

\begin{document}

\title{Beyond the Brim of the Hat: Kinematics of Globular Clusters out to Large Radius in the Sombrero Galaxy}

\author{Jessica L. Dowell and Katherine L. Rhode}
\affil{Department of Astronomy, Indiana University, Swain West 319, 727 East Third
  Street, Bloomington, IN 47405-7105; jlwind@astro.indiana.edu, rhode@astro.indiana.edu} 

\author{Terry J. Bridges}
\affil{Department of Physics, Engineering Physics, and Astronomy, QueenÕs University, Kingston, ON K7L 3N6, Canada; tjb@astro.queensu.ca}

\author{Stephen E. Zepf}
\affil{Department of Physics and Astronomy, Michigan State University, East Lansing, MI 48824; zepf@pa.msu.edu}

\author{Karl Gebhardt}
\affil{Department of Astronomy, University of Texas at Austin, 1 University Station C1400, Austin, TX 78712; gebhardt@astro.as.utexas.edu}

\author{Ken C. Freeman}
\affil{Research School of Astronomy and Astrophysics, Australian National University, Mount Stromlo Observatory,
Weston Creek, ACT 2611, Australia; kcf@mso.anu.edu.au}

\and

\author{Elizabeth Wylie de Boer}
\affil{Research School of Astronomy and Astrophysics, Australian National University, Mount Stromlo Observatory,
Weston Creek, ACT 2611, Australia; ewylie@mso.anu.edu.au}

\begin{abstract}
\label{section:abstract}
We have obtained radial velocity measurements for 51 new globular clusters around the Sombrero galaxy.  These measurements were obtained using spectroscopic observations from the AAOmega spectrograph on the Anglo-Australian Telescope and the Hydra spectrograph at WIYN.  Combined with our own past measurements and velocity measurements obtained from the literature we have constructed a large database of radial velocities that contains a total of 360 confirmed globular clusters.  Previous studies' analyses of the kinematics and mass profile of the Sombrero globular cluster system have been constrained to the inner $\sim$9\arcmin~($\sim$24 kpc or $\sim$5$R_e$), but our new measurements have increased the radial coverage of the data, allowing us to determine the kinematic properties of M104 out to $\sim$15\arcmin~($\sim$41 kpc or $\sim$9$R_e$).  We use our set of radial velocities to study the GC system kinematics and to determine the mass profile and V-band mass-to-light profile of the galaxy.  We find that $M/L_V$ increases from 4.5 at the center to a value of 20.9 at 41 kpc ($\sim$9$R_e$ or 15\arcmin), which implies that the dark matter halo extends to the edge of our available data set.  We compare our mass profile at 20 kpc~($\sim$4$R_e$ or $\sim$7.4\arcmin) to the mass computed from x-ray data and find good agreement.  We also use our data to look for rotation in the globular cluster system as a whole, as well as in the red and blue subpopulations.  We find no evidence for significant rotation in any of these samples.
\end{abstract}


\section{Introduction}
\label{section:introduction}
While the details of galaxy formation are not yet well understood, the current paradigm suggests that dark matter (DM) halos play a critical role in the process.  In these halos, baryonic matter collects and cools to form stars and galaxies, and it is believed that the subsequent merging of these halos and their contents leads to the formation of more massive galaxies.  Thus, understanding the structure of DM halos is fundamentally important for testing galaxy formation models and cosmological theories.  One way to examine the DM halo of a galaxy is to analyze its mass profile out to large radii.  For gas-rich galaxies such as spirals, this can be done by examining the kinematics of the stars and the neutral hydrogen gas.  However this type of analysis is much more difficult for early-type galaxies since they lack these easily observed dynamical tracers.  Globular cluster systems provide an excellent set of alternative tracers for exploring the outer regions of early-type galaxies.  Globular clusters (GCs) are luminous, compact collections of stars that are billions of years old and formed during the early stages of galaxy formation \citep{Ashman98, Brodie06}.  They have been identified in photometric studies out to 10 to 15 effective radii (e.g. \citet[hereafter RZ04]{Rhode04}, \citet{Harris09}, \& \citet{Dirsch03}) and, therefore, serve as excellent probes of the formation and merger history of their host galaxies \citep{Brodie06}.  Unfortunately, due to observational constraints, few GC systems have large numbers (more than 100-200) of spectroscopic radial velocity measurements necessary for these types of kinematic studies.  Some of the GC systems with the largest number of measured radial velocities include those around massive elliptical galaxies, such as NGC4472 \citep{Cote03}, NGC1399 \citep{KP98, Dirsch04, Schuberth10}, and M87 \citep{Cote01, Hanes01, Strader11} as well as the S0 galaxy NGC5128 \citep{Peng04, Woodley07}.

M104, otherwise known as the Sombrero Galaxy or NGC 4594, is an isolated edge-on Sa/S0 galaxy located at a distance of 9.8 Mpc \citep{Tonry01} with an effective radius of 1.7\arcmin~(4.6 kpc) \citep{Kormendy89}.  Several photometric studies have been made of the GC system of M104.  RZ04 performed a wide-field photometric study of the GC system of the galaxy in $B,V,$ \& $R$, and detected GC candidates out to 25\arcmin~(15$R_e$ or 68 kpc).  They also found that the GC system of the Sombrero contains roughly 1900 clusters with a de Vaucouleurs law radial distribution that extends to 19\arcmin~(11$R_e$ or 51 kpc), where extent is defined as the radius where the surface density of GCs is consistent with zero within the estimated measurement errors.  \citet{Larsen01}, \citet{Spitler06}, and \citet{Harris10} performed HST photometry on the more crowded central regions of M104 in order to detect GC candidates closer to the center of the galaxy.  All of these photometric studies found that the GC system of M104, like those of many giant galaxies, exhibits a bimodal color distribution which is assumed to correspond to a metal-rich red subpopulation and a metal-poor blue subpopulation \citep{Gebhardt99, Kundu01, Rhode04}.  Examining the kinematics of GCs in these two sub-populations can test whether or not they formed during two distinct phases of galaxy formation.

In addition to these photometric studies, several groups have performed spectroscopic observations of the M104 GC system \citep{Bridges97, Bridges07, Larsen02, Deimos11}. \citet[hereafter B07]{Bridges07} performed a relatively wide-field kinematic study using spectroscopically measured radial velocities for 108 GCs out to 20\arcmin~(12$R_e$ or $\sim$54 kpc) with the 2dF spectrograph on the Anglo-Australian Telescope (AAT). They found that the $M/L_V$ ratio of the galaxy increases with distance from the center to $\sim$12 at 9.5\arcmin~(6$R_e$ or $\sim$25 kpc), which provides tentative support for the presence of a DM halo around M104. In addition, they found no evidence of global rotation in the GC system or in the red and blue sub-populations. However, the limited number of GCs at large radii in this study makes the results in the outer regions uncertain.  The most recent spectroscopic study by \citet{Deimos11} consists of a large number of clusters (over 200); however, they only observe GCs out to a distance of about 27 kpc ($\sim$6$R_e$ or $\sim$10\arcmin) from the center of the galaxy.  In addition, they did not perform a kinematic analysis of their sample.  They did, however, confirm the metallicity bimodality of the GC system detected in the earlier photometric studies (peaks at [Fe/H] = --1.4 and [Fe/H] = --0.6).  In order to acquire a more complete understanding of the dynamical properties of the galaxy, it is crucial to obtain both large numbers of velocity measurements and measurements which provide significant spatial coverage. 

We have obtained new spectroscopic observations of M104 GCs using the AAOmega spectrograph on the 3.9m AAT and the Hydra spectrograph on the WIYN 3.5m telescope.  From these data we obtained 51 new GC velocities and we combine these new measurements with data from the literature to create a sample of 360 confirmed M104 GCs with reliable radial velocity measurements that include objects out to 24\arcmin~($\sim$14.1$R_e$ or $\sim$64.9 kpc) in galactocentric distance.  This is the largest sample of radial velocity measurements used for a kinematic study for the Sombrero to date. Using this sample, we were able to study the kinematics and mass profile of M104 to a larger radial extent than previous studies.

The paper is organized as follows:  Section 2 covers the acquisition and processing of the data, while the methods used to obtain radial velocity measurements for our target GC candidates are discussed in \S3.  Section 4 provides an analysis of the rotation within the GC system and the determination of the mass profile.  Section 5 provides a discussion of our results compared to kinematic studies of other mass tracers in M104 and other galaxies.  Finally, \S6 summarizes our findings.

\section{Observations \& Data Reduction}

\subsection{Observations}
\label{section:observations}

\subsubsection{AAOmega Observations}
\label{section:aaomega}

Spectroscopic observations were acquired for this study during two observing runs.  The first spectra were taken with the Anglo-Australian telescope using the AAOmega multi-object spectrograph in May 2009.  AAOmega is fed by the Two-Degree Field (2dF) fiber positioner which contains 392 fibers.  The wide field of view of AAOmega is well suited for our observations since the GC system of M104 extends to at least 19\arcmin~($\sim$11$R_e$ or $\sim$51 kpc) from the galaxy center \citep{Rhode04}. For each fiber the dual-beam spectrograph produces both a red and a blue spectrum.  For our observations, the red arm of the spectrograph was configured with the 1000I VPH grating centered at 8580 \AA, resulting in spectra with a dispersion of $\sim$0.54 \AA/pixel and a wavelength range of $\sim$8015-9120 \AA. The blue arm of the spectrograph was configured with the 580V VPH grating centered on a wavelength of 4750 \AA, yielding spectra that have a central dispersion of $\sim$1.04 \AA/pixel and a wavelength range of $\sim$3680-5800 \AA.  It should be noted that for the purpose of our analysis the red and the blue AAOmega spectra were treated as two independent data sets.  

Targets for our observations were selected from the photometric GC system study of RZ04, in which 1748 candidate GCs around M104 were identified using wide-field BVR images.  From the RZ04 list we selected $\sim$500 candidates without measured velocities and with V magnitudes between 19.0 and 22.0.  From this magnitude-limited list we created two spectrograph configurations to be observed over three nights with AAOmega.  Selection of GC candidates for the two configurations was weighted by magnitude, with brighter objects receiving higher priority.  Due to inclement weather we were only able to observe one of the two configurations during the run.  Seeing during both nights of observations was poor and ranged from $\sim$2-3\arcsec.  Our resulting data set is comprised of 13 1800-second exposures that include a total of 268 GC candidates.  Internal quartz lamp flats and arc lamp spectra were also taken for calibration purposes. 

\subsubsection{Hydra Observations}
\label{section:hydra}

Additional GC candidates were observed using the Hydra spectrograph on the WIYN 3.5m telescope\footnote{The WIYN Observatory is a joint facility of the University of Wisconsin-Madison, Indiana University, Yale University, and the National Optical Astronomy Observatory.} at Kitt Peak National Observatory (KPNO)\footnote{Kitt Peak National Observatory, part of the National Optical Astronomy Observatory is operated by the Association of Universities for Research in Astronomy (AURA) under a cooperative agreement with the National Science Foundation.} in February 2011.  Hydra is a multi-object spectrograph with $\sim$85 available fibers and a one degree field of view.  Targets for the Hydra observations were selected from the remaining unobserved objects in the RZ04 list described above.  The throughput of the Hydra spectrograph is lower than that of AAOmega, so we further restricted our target selection to only those objects with V magnitudes brighter than 20.5.  Our resulting target list contained 72 GC candidates.  We observed a total of 12.5 hours on a single fiber configuration which contained a total of 48 GC targets.  Due to a combination of high winds and high humidity, the seeing was relatively poor for WIYN and ranged from 1\arcsec~to 2\arcsec~throughout our observations.  Dome flats, dark frames, and arc spectra were also obtained during the run.  For the observations, Hydra was configured with the red fiber bundle and the 600@10.1 grating centered at a wavelength of 5300 \AA, providing a spectral range that extends from 3880 to 6710 \AA.  This is similar to the spectral coverage of the blue arm of the AAOmega data.  To improve the signal-to-noise (S/N) of our spectra we also binned the pixels of the STA1 CCD two by two.    

\subsection{Reductions}
\label{section:reductions}

\subsubsection{AAOmega}
The red and blue AAOmega spectra were reduced using the 2dfdr pipeline software \citep{Croom05}.  The spectra were first flat-fielded and then wavelength-calibrated using the arc lamp spectra.  Next, a throughput calibration was performed using night sky lines, and the sky was subtracted using dedicated sky fibers.  Finally, the individual spectra were extracted using a weighted extraction routine, and the object spectra were stacked to produce a final combined set of spectra.  

Upon examination of the reduced data, we discovered that some fibers in each of the blue exposures exhibited poor calibration of the wavelength zero-point, with errors $\sim$1-3 \AA.  After further investigation we found that this problem affected a different set of fibers in each exposure, such that a given fiber might have bad calibration in only one or two of the 13 exposures.  To circumvent this problem, we restacked the blue spectra manually using the IRAF\footnote{IRAF is distributed by the National Optical Astronomy Observatories, which are operated by the Association of Universities for Research in Astronomy, Inc., under cooperative agreement with the National Science Foundation.} task SCOMBINE, one fiber at a time, excluding any spectra with wavelength calibration errors greater than 1 \AA.  

Finally, to prepare the spectra for cross-correlation, we trimmed $\sim$10 \AA~from the ends of each spectrum to remove any potential edge artifacts, and then continuum-subtracted the data using a Legendre polynomial fit (eighth-order for the blue spectra and ninth-order for the red).  

\subsubsection{Hydra}
The Hydra spectra were reduced using the IRAF DOHYDRA spectral reduction package.  Before running DOHYDRA, the flat field, arc lamp, and object spectra were bias-subtracted and corrected for dark current using CCDPROC.  Next, DOHYDRA was used to extract the individual spectra and perform the throughput correction.  Wavelength calibration of the object spectra was then performed using the arc lamp spectra.  The final root-mean-square error of the best-fit wavelength solution was less than 5\% of the dispersion of the spectra.  Finally, the spectra were sky subtracted.  Once the spectra were reduced they were scaled to the flux level of the brightest observation, cosmic ray cleaned, and stacked using SCOMBINE.  To remove any edge effects $\sim$10 \AA~was clipped from the two ends of each spectrum.  Finally, the continuum for each spectrum was fit with an eighth-order Legendre polynomial, and subtracted to produce the final data set.

\section{Analysis} 
\label{section:analysis}

\subsection{Measuring the Globular Cluster Velocities}
\label{section:velocities}
Heliocentric radial velocities for our target objects were measured using the IRAF FXCOR task, which performs a Fourier cross-correlation of each object spectrum against a set of template spectra.  Our template spectra were generated using the GALAXEV stellar population synthesis code \citep{Bruzual03}.  We chose the GALAXEV model templates since this allowed us to use one consistent set of templates for all of our spectra which cover a wide wavelength range from 3680 \AA~to 9120 \AA.  Six templates were created, each assuming a single instantaneous burst of star formation with a Salpeter initial mass function and a range of metallicities from an [Fe/H] of --2.25 to +0.56.  These values fully encompass the metallicity range for M104 GCs found by \citet{Deimos11}.  

Each run of the model was allowed to evolve for 12 Gyr, which is consistent with the 10-15 Gyr age spread for M104 GCs determined by \citet{Larsen02}.  We ran additional tests using models with a range of ages spanning a few Gyr around our chosen value.  However, at a given metallicity these small changes in the age of the models did not produce different cross-correlation results, and consequently we chose to use a single age for our final templates.    

Once the spectra were cross-correlated with the GALAXEV templates, we had six velocity measurements and associated FXCOR uncertainties for every object in each of the three data sets.  We then performed several cuts on each set of measurements to eliminate spurious values and weak detections from our sample.  For bright target objects with high S/N (roughly 16\% of the targets), the cross-correlation routine measures roughly the same velocity value for each of the six model templates.  However, as the S/N of the target object spectra decreases, one or more of the templates may fail to produce a strong cross-correlation peak, resulting in velocity measurements that are wildly different from the rest of the measurements for that object.  Keeping these spurious measurements would skew the calculation of the final mean velocity for these objects.  Consequently, our first step was to examine the self-consistency of the six velocity measurements for each object by comparing each measurement with the median value for the ensemble.  We used standard error propagation to compute the uncertainty on the median using the errors on the individual measurements from FXCOR.  Those measurements that fell outside 3-$\sigma$ from the median value were rejected.  

With a self-consistent set of measurements for each object, we next determined which measurements were too weak to be considered reliable. The most common way to do this is to use the Tonry-Davis $R$ (TDR) coefficient \citep{TonryDavis} which is defined as 
\begin{equation}
\label{eq:tdr}
R = \frac{h}{\sqrt{2}\sigma_a},
\end{equation}
where $h$ is the height of the true peak of the cross correlation function, and $\sqrt{2}\sigma_a$ is the average height of all the peaks in the cross-correlation function.  Selection of the $R$ value cutoff is a balance between maximizing the final sample size while also ensuring the reliability of the final velocity measurements.  To determine our cutoff value we first cross-correlated nine of the sky spectra (that were observed for calibration purposes with our data) with our six model templates using the same FXCOR parameters that we used for the object spectra.   This allowed us to estimate the values of $R$ that could be produced by cross-correlating our templates with noise.  We established a preliminary cut at an $R$ value of 3.6 based on the mean $R$ from the sky cross-correlations.  We were then able to take advantage of the fact that we had dual AAOmega spectra to refine this limit.  We compared the mean velocity for each object from the red AAOmega spectra to its corresponding mean velocity computed in the blue data.  For a given object, the two data sets should produce the same velocity.  Consequently, we incrementally increased the $R$ value cut until the majority of the objects with highly discrepant, i.e., $\gg$3-$\sigma$ or 400 km s$^{-1}$, velocities between the red and blue data were eliminated from the sample.  Using this, we determined the optimal $R$ cut to be at a value of 4.25.  Any measurements below this threshold were removed from the subsequent analysis.

Finally, we performed a rough velocity cut, and for each object, we removed velocity measurements less than --500 $km s^{-1}$ or greater than 15,000 $km s^{-1}$.  The lower limit of --500 $km s^{-1}$ encompasses the expected velocities of most nearby Galactic stars.  For objects with low S/N, a pattern of strong, high velocity peaks above $\sim$15,000 $km s^{-1}$ began to appear in some of the cross correlation functions.  These are unlikely to be real since the strength of these peaks is inconsistent with the low S/N of the data and they occur at a frequency much higher than expected for genuine high-redshift objects.  As a result, we chose to reject these individual measurements as unreliable.  For the objects with one or more velocity measurements remaining after this series of cuts, the individual measurements were combined using an uncertainty-weighted mean to produce a final velocity measurement for that object.  On average, four of the six velocity measurements for each object survived the cuts and were used in this final averaged velocity.  The final uncertainties for the mean velocity for each object were computed by propagating the FXCOR errors from the individual measurements included in the mean.  For reference, the average velocity uncertainty for an object is $\approx$19 km s$^{-1}$.  After this series of cuts, 18 objects remained from the 48 measured with Hydra.  48 objects remained from the blue AAOmega data set, and 117 objects remained in the red AAOmega data set.

\begin{figure}
\plotone{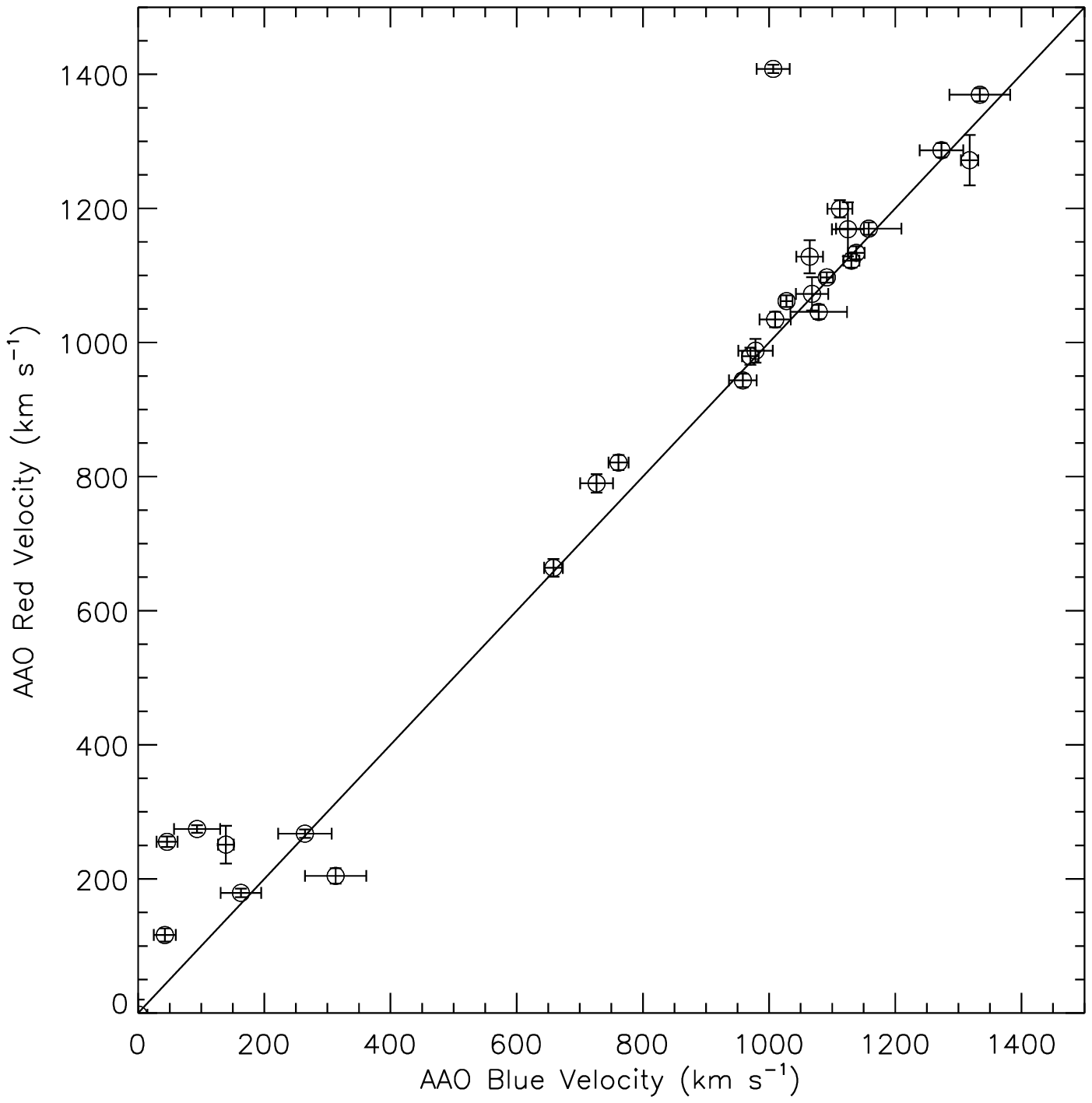}
\caption{ AAOmega red velocity versus AAOmega blue velocity for the 30 matching objects that pass the velocity selection criteria.  The solid black line indicates a 1:1 velocity ratio.  The mean velocity difference is 45.5 $\pm$ 92.4 $km s^{-1}$, indicating that there is good overall agreement between the velocities determined from the two data sets. We find no significant systematic offset, which suggests that our cross correlation methods are self consistent.  }  
\label{fig:rbcomparison}
\end{figure}

There are 30 objects in common between the final red and blue AAOmega samples.  Figure \ref{fig:rbcomparison} shows the comparison between the red and the blue velocities for each of these objects.  The mean difference between the red and blue velocities is 45.5 km s$^{-1}$with a dispersion of 92.4 $km s^{-1}$ which suggests that our method of measuring velocities is self-consistent across our different data sets.  One object in Figure \ref{fig:rbcomparison} (RZ 4093) exhibits a large velocity difference of 401.5 $km s^{-1}$.  The velocity measurement of this outlier is much weaker in the blue data set (mean $R$ of $\sim$4.7) than in the red, where the cross-correlation is strong (mean $R$ of $\sim$8.7).  As a result, we adopt the velocity from the red AAOmega data for this particular object.  In addition to this outlier,  objects below $\sim$500 km s$^{-1}$ show a larger scatter around the one-to-one line, and are likely Galactic stars (see \S\ref{section:outliers}).  By design, the cross-correlation templates are best suited for detecting GCs, and as a result, the velocities for these objects are not as well determined.  If we exclude these objects and RZ 4093 we find that the mean difference between the red and blue objects is reduced to 18.2 km s$^{-1}$ with a dispersion of 33.7 km s$^{-1}$.  This difference is comparable to the mean FXCOR uncertainties of $\approx$12 km s$^{-1}$ for the red data set and $\approx$28 km s$^{-1}$ for the blue data set.  Furthermore, if we examine the objects in this reduced sample individually, we find that that  35\% have blue measurements within 1-$\sigma$ of the red value and 65\% have red measurements within 1-$\sigma$ of the blue value.  This suggests that although the overall agreement between the red and blue data sets is good, the uncertainties determined by FXCOR for the red spectra may be underestimated.  For the analysis of our data, we adopted a weighted average of the red and blue velocities for these 30 objects.  In addition to the objects repeated within the AAOmega data sets, there is also one object, RZ 2832, from the Hydra data which is repeated in the AAOmega data.  As with the red and blue data, we use the weighted average of the available velocities as the final velocity for this object.  Our final observed sample of velocities consists of 152 unique objects.

\subsection{Published Velocities}
\label{section:literature}
In addition to our own velocity measurements, we assembled an additional sample of velocities using the results from previous work in the literature.  First, we include 46 velocity measurements obtained by \citet{Bridges97} using the William Herschel Telescope and the Low Dispersion Survey Spectrograph. An additional 16 objects were obtained from Keck Low Resolution Imaging Spectrometer observations of \citet{Larsen02}.  This excludes the object identified as H2-27 in the \citet{Larsen02} list since the authors declare the velocity measurement to be unreliable.  B07 determined velocities for 170 objects with the 2dF spectrograph at the AAT and with the Hydra spectrograph on WIYN.  Finally, \citet{Deimos11} measured 259 velocities for GC candidates using Keck and the DEep Imaging Multi-Object Spectrograph.  We cross-matched the objects in each of these lists and found repeated velocity measurements for a total of 38 objects, including two objects that were repeated more than once.  As with the repeats in our own observations, we adopted the uncertainty-weighted mean of all available velocity measurements as the velocity for these objects.  The resulting literature sample consists of 450 unique objects.

Each of these studies used a different set of criteria to separate true GCs from contaminating objects such as foreground stars. Therefore, in order to implement a consistent method to separate true GCs from contaminating objects we have included velocities for all science objects measured in the literature, including those previously designated as contaminants.  

\subsection{Removing Duplicates}
\label{section:deduplication}

\begin{figure}
\plotone{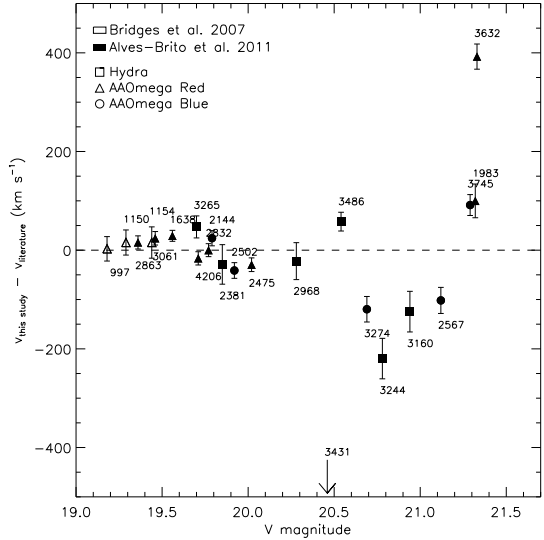}
\caption{The difference between our velocity measurements and values from previous studies plotted as a function of V magnitude.  Open symbols show results compared to values from \citet{Bridges07}, and filled symbols are a comparison to values from \citet{Deimos11}.  The various symbol shapes indicate from which of our data sets each velocity was measured: squares indicate velocities from the Hydra data, triangles indicate velocities from the red AAOmega data, and circles represent velocities from the blue AAOmega data.  We find good agreement between our values and the literature, overall, with an expected increase in scatter toward fainter magnitudes.}  
\label{fig:comparison}
\end{figure}

There are 23 objects in our observed sample that also match objects in the literature.  These repeated velocity determinations provide a way to independently check our cross-correlations, and give us an estimate of the systematic errors associated with our velocity measurements.  Figure \ref{fig:comparison} shows the difference between our velocity measurements and the literature values plotted against V magnitude for each of the 23 objects.  Overall we find good agreement between the repeated measurements.  The mean velocity difference for the 23 repeated objects is --23.1 km s$^{-1}$ with a dispersion in the values of 174.9 $km s^{-1}$, which is consistent with zero, given their mean velocity uncertainty of $\approx$26 km s$^{-1}$.  The increased spread seen in the data points at fainter magnitudes is expected since the spectra for these objects have decreased S/N, and as a result, have a greater uncertainty in their cross correlations.  Indeed, if we consider only objects brighter than a V-band magnitude of 20, we find much better agreement, and the mean velocity difference becomes 7.3 km s$^{-1}$ with a dispersion of 25.5 km s$^{-1}$.  There is one notable outlier, RZ 3431.  Even though it has a moderate V magnitude of 20.5, for this object we measure a velocity that is over 600 $km s^{-1}$ lower than the value reported by \citet{Deimos11}.  Examination of our cross-correlation results for this object reveal that our velocity measurement is consistent across all six templates, and has a TDR value greater than the cutoff for four of our six measurements.  As a result, even though the \citet{Deimos11} velocity has a lower uncertainty, we adopt our own velocity for this target.  If we exclude this outlier from the calculation of the mean difference, we find a value of 5.1 km s$^{-1}$ with a dispersion of 114.0 $km s^{-1}$ (for all magnitudes), which shows even stronger agreement between our measurements and the literature.
Once again, in the subsequent analysis we adopt the weighted average of the velocities for the repeated objects.  After removing these duplicate measurements, our final data set contains velocities for a total of 579 unique GC candidates.  Table \ref{table:gcdata} lists complete data for all of the objects for which we were able to measure a radial velocity.  Whenever possible, the objects are identified using the sequence numbers assigned by RZ04 (given as RZ\# in the table).  In those cases where an object did not have an RZ04 identifier, we adopted the identifier number from the literature in the following order of preference; \citet{Spitler06} (S\#), \citet{Larsen02} {C-\# or H-\#}, and \citet{Bridges97} (1-\# or 2-\#).   

\subsection{Contaminant Rejection}
\label{section:outliers}

\begin{figure}
\plotone{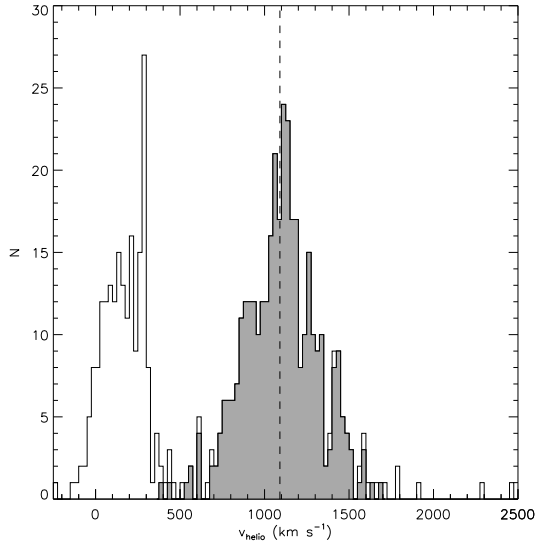}
\caption{Velocity histogram of all 577 objects in our combined sample with velocities less than 2500 $km s^{-1}$ (there are two objects in our final sample with velocities greater than 2500 $km s^{-1}$ that are not shown).  Objects with velocities within the large peak centered over 1091 $km s^{-1}$ are members of the M104 GC system, and objects with velocities in the second large peak at 150 $km s^{-1}$ are Galactic foreground stars.  The tails of the two peaks overlap near $\sim$500 $km s^{-1}$.  The shaded region shows the final GC sample selected using the TME technique of \citet{Schuberth10, Schuberth12}.} 
\label{fig:vhist}
\end{figure}

\begin{figure}
\plotone{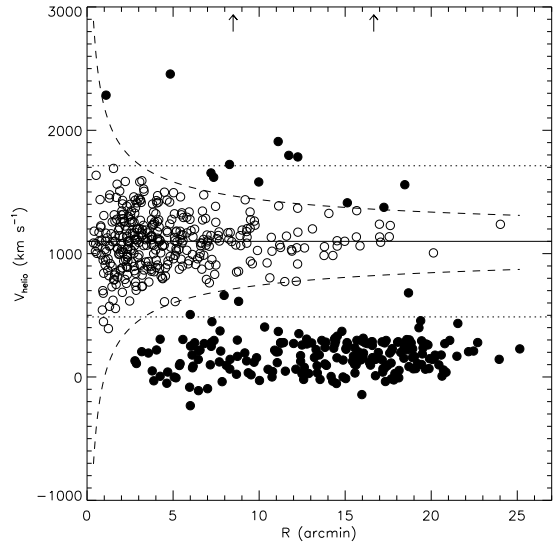}
\caption{Results of the contaminant rejection algorithm.  The filled points are the objects rejected as contaminants, and open points represent the objects in our final sample of confirmed GCs.  The dashed line indicates the boundary established by the \citet{Schuberth10, Schuberth12} rejection routine used to select our final GC sample.  Finally, the solid black line indicates the mean velocity of 1099.5 $\pm$ 204.2 $km s^{-1}$ for the GCs computed by fitting a double Gaussian to the velocity data.  The dotted lines demonstrate the result of using a flat velocity cutoff at 3-$\sigma$ above and below this mean value to separate GCs from contaminants.  Two rejected contaminants have velocities greater 3000 $km s^{-1}$, and are indicated using arrows.} 
\label{fig:rejection}
\end{figure}

The photometric studies of globular cluster candidates that we used to select targets for our spectroscopic follow-up utilized broadband magnitudes and colors to select GCs around the Sombrero galaxy, e.g., RZ04.  As a result, there is some contamination inherent in the data set which is primarily due to foreground stars and compact background galaxies that exhibit magnitudes and colors similar to GCs.  Figure \ref{fig:vhist} shows the velocity histogram for all objects in our sample.  The cluster of velocities near the systemic velocity of M104 at 1091 $km s^{-1}$ \citep{Tully08} are GCs, while the objects at lower velocities are Galactic stars.  Previous studies were able to take advantage of a large gap in the velocity histogram near $\sim$500 $km s^{-1}$ to separate GCs from the contaminants \citep{Bridges07}.  In our much larger sample, this break is less apparent, however we can still apply a similar velocity cut to our data.  We fitted the velocity histogram in Figure \ref{fig:vhist} using a double Gaussian function.  From the resulting best fit we determine that the mean velocity of the GC peak in our sample is located at 1099.5 $\pm$ 204.2 $km s^{-1}$.  Objects with velocities within 3-$\sigma$ (indicated by the dotted lines in Figure \ref{fig:rejection}) of this value are most likely M104 GCs since the probability of finding a star with a velocity greater than 486 km s$^{-1}$ is only 1.0\%.  368 objects in our sample have velocities that fall within this 3-$\sigma$ range.

\begin{figure}
\plotone{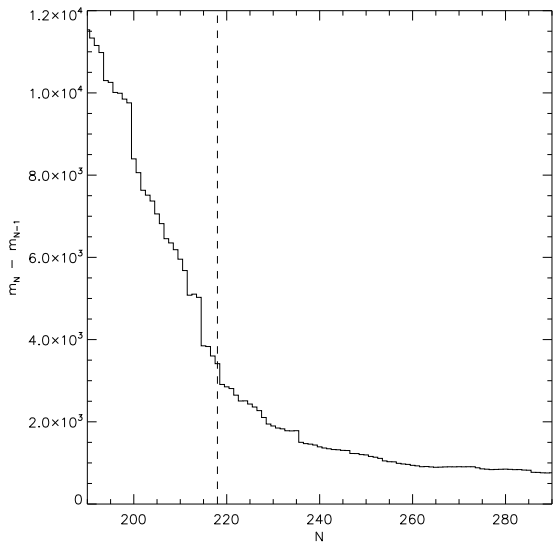}
\caption{Difference between the $m_N$ and $m_{N-1}$ values computed from Equation \ref{eq:tme} plotted against N for each step of the \citet{Schuberth10, Schuberth12} rejection algorithm.  The vertical dotted line indicates our chosen rejection limit, which is the point at which the differences begin to converge. Using this cut we reject a total of 219 contaminating objects.} 
\label{fig:rejectindex}
\end{figure}

Another method that can be used to identify GCs is the outlier rejection method detailed in \citet{Schuberth10} and \citet{Schuberth12}, which utilizes the tracer mass estimator (TME) derived in \citet{Evans03}.  One advantage of this method is that the selection is based on the projected distance of each object relative to the center of the galaxy, in addition to the radial velocity of each object.  To use this method, we first compute the value $m_N$, which is defined as
\begin{equation}
\label{eq:tme}
m_N = \frac{1}{N}\sum\limits_{i=1}^N v_i^2R_i\mbox{,}
\end{equation}
where $m_N$ is a proxy for the TME, N is the number of GC candidates, $v_i$ is the velocity of the candidate relative to the galaxy (1091 $km s^{-1}$ \citep{Tully08}), and $R_i$ is its projected radial distance from the galaxy center.  For computing the value of $R_i$, we adopt a value for the galaxy center from NED at an RA of $12^h39^m59.43^s$ and a declination of $-11^{\circ}37'23.0"$.  Next, the object with the largest value of $v^2R$ is removed, and Equation \ref{eq:tme} is recomputed using the remaining sample. Figure \ref{fig:rejectindex} shows the difference between the $m_N$ and $m_{N-1}$ plotted against N at each step.  Objects that belong to the M104 system should trace the same overall mass, and removal of the contaminating objects will cause the values of $m_N$ to converge. Consequently, the cutoff between contaminants and GCs in this figure will appear as a flattening of the curve.  We selected the last large jump in the value of $m_N$ -- $m_{N-1}$ at 218 as our cutoff value.  Our chosen cutoff between true GCs and contaminants is shown by the vertical dashed line in Figure \ref{fig:rejectindex}. This selection represents a tradeoff between removing the contaminants and maximizing the GCs in our final sample.  The next largest jumps in Figure \ref{fig:rejectindex} above and below our chosen cutoff are at 214 and 227, which suggests that the uncertainty in the number of GCs in our final sample is on the order of ten GCs.  

Figure \ref{fig:rejection} shows the division of contaminants from GCs resulting from this method on a plot of relative velocity vs. projected galactocentric distance.  The boundary computed with the TME method is shown by the dashed curve.  Rejected objects are indicated by filled circles, and confirmed GCs are indicated by open circles.  For illustrative purposes, the flat velocity cut discussed above is shown using dotted lines.  Using our chosen cutoff from the TME method, we eliminate a total of 219 contaminating objects from our sample.  Of the rejected objects, 205 are low velocity objects that are most likely foreground stars associated with the Galaxy.  The remaining 14 high velocity objects are most likely a combination of background galaxies and spurious measurements of low S/N objects.  The 360 confirmed GCs in our list extend to a distance of 24\arcmin~($\sim$14.1$R_e$ or $\sim$64.9 kpc) from the center of the galaxy and span a range of V magnitude from 18.79 to 23.72 and a B--V color range from 0.42 to 1.20.  We use this sample of GCs for the remainder of our analysis.  

\section{Results}
\label{section:rotation}

\subsection{Distribution}
\begin{figure}
\plotone{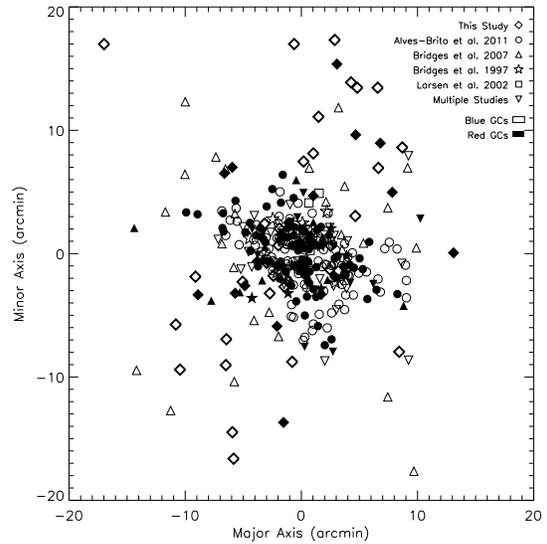}
\caption{Spatial distribution of the spectroscopically confirmed GC sample.  The center of the Sombrero galaxy is located at the center of the plot at [0,0].  The different shaped symbols indicate the source of the velocity measurement for each cluster.  GCs with final velocities that were averaged from multiple sources are denoted by upside-down triangles.  Our new data from AAOmega and Hydra, indicated by diamond symbols, doubles the number of GC velocities beyond 10\arcmin~(6$R_e$ or $\sim$27 kpc), and indicates a significant improvement in radial coverage over previous studies.  Filled and open symbols indicate whether a particular cluster belongs to the red or blue subpopulation, respectively.} 
\label{fig:spatial}
\end{figure}

Figure \ref{fig:spatial} shows the spatial distribution of our confirmed GCs around the center of the Sombrero galaxy, which is located at the origin of the plot.  The different symbols indicate the source of the velocity measurement for each object.  GCs with final velocities that were averaged from multiple sources are denoted by upside-down triangles.  Notably, our new velocity measurements obtained with AAOmega help to significantly improve the coverage of the data at radii greater than $\sim$10\arcmin~(6$R_e$ or $\sim$27 kpc).  B07, who performed the most recent kinematic study of the M104 GC system, measured velocities for 16 GCs beyond 10\arcmin.   Our new data set includes an additional 19 GCs outside of 10\arcmin, which more than doubles the number of GC velocities beyond this radius.  The shading of the symbols in Figure \ref{fig:spatial} indicates whether the GC belongs to the red or blue subpopulation, where the division between red and blue is designated as a B--R color of 1.3 \citep{Rhode04}.  We identify 210 objects as blue GCs and 149 objects as red GCs.  Note that there is one object in our final GC sample, GC \#1-33, that does not have an available B--R color, and so although it is included in the analysis of the full GC sample, it is excluded from our analysis of the subpopulations.  GC \#1-33 was classified as a star by \citet{Bridges97} and was not observed photometrically by RZ04.  As a result, its photometry was unreported in the literature.  We measure its velocity as 448 $\pm$ 141 $km s^{-1}$, which is low for M104 GCs; however, because of its low projected distance from the center of the galaxy, it survives our contaminant rejection routine, and we therefore classify it as a GC in our sample.  An initial visual inspection of Figure \ref{fig:spatial} appears to indicate that there is a difference in the distribution of red and blue GCs in our sample.  Indeed, beyond a radius of $\sim$10\arcmin~(6$R_e$ or $\sim$27 kpc), there is a greater proportion of blue GCs to red GCs with 27 blue GCS and 8 red GCs.  RZ04 observed a small radial color gradient in their 90\% complete sample of M104 GCs with a slope of $\Delta(B-R) / \Delta(r) =$ --0.003$\pm$0.001, which would be consistent with a changing ratio of blue to red objects at large galactocentric radii.  However, a two-sample Kolmogorov-Smirnov test on the full sample gives  a p-value of 0.03 which indicates that the difference between the red and blue distributions is not significant, and we cannot rule out the possibility that the red and blue GCs are from the same parent distribution.  Furthermore, a K-S test including only those GCs outside 10\arcmin~results in a p-value of 0.53, which further suggests that the two subpopulations within our sample do not have different radial distributions.

\subsection{Rotation}
\begin{figure}
\plotone{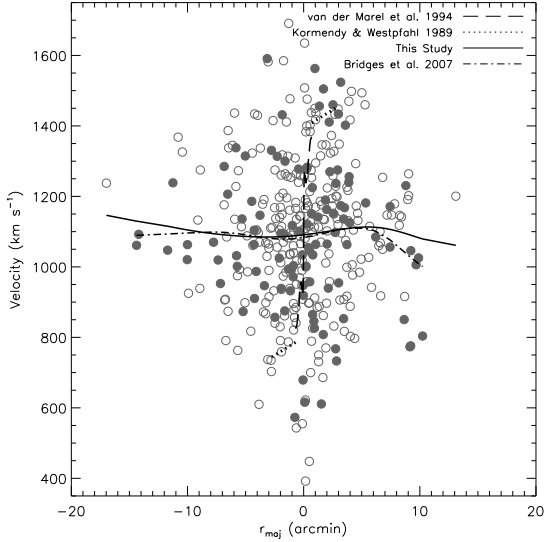}
\caption{Radial velocity as a function of distance along the galaxy major axis for our GC sample.  Our sample of GCs is indicated by circles.  The solid black line shows the result of smoothing all of the data points with a Gaussian with a standard deviation of 3\arcmin~($\sim$2$R_e$ or $\sim$8 kpc).  The dash-dotted line shows the result obtained by B07 using the same method for the GCs in their data set, which are indicated by the filled the circles.  The dashed lines and dotted lines show the rotation curves of stars and gas for the Sombrero from \citet{Vandermarel94} and \citet{Kormendy89}, respectively.  In contrast to the strong rotation of the stars and gas, we see no significant rotation in the GC system.} 
\label{fig:rotation1}
\end{figure}

We examined the radial velocities of our GCs as a function of projected distance along the major axis of M104.  The resulting plot is shown in Figure \ref{fig:rotation1}.  The solid line shows the result of performing smoothing to the data with a sliding average using a Gaussian weighting with standard deviation of 3\arcmin~($\sim$2$R_e$ or $\sim$8 kpc).  For comparison, the dash-dotted line shows the result of this treatment determined by B07.  Like B07, we find no evidence for obvious rotation in the GC system.  This is in sharp contrast to the strong rotation seen in the stellar rotation curve of the galaxy \citep{Vandermarel94} and in the HII emission and HI gas rotation curve from \citet{Kormendy89}, shown by the dashed and dotted lines, respectively.  

\begin{figure}
\plotone{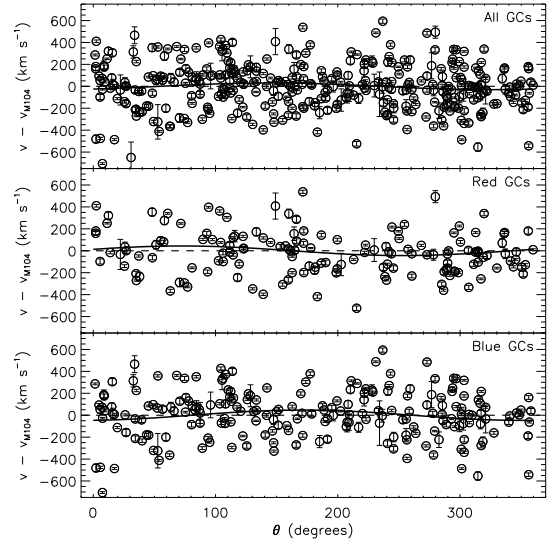}
\caption{Relative GC velocity as a function of position angle around the galaxy.  The top panel shows the results for the full GC sample, the center shows the results for the GCs in the red subpopulation, and the bottom panel shows the results for the blue GC subpopulation.  For the systemic velocity of M104, we adopt the robust, biweight mean velocity of the full GC sample of 1097.3 $km s^{-1}$.  The solid black lines in each panel show the best fit curve to Equation \ref{eq:rotation} determined by a nonlinear least-squares fit to each of the data sets.  We find no indication of significant rotation in the full GC system or in either subpopulation using this method.} 
\label{fig:rotation2}
\end{figure}

We performed a closer examination of rotation by looking at radial velocity as a function of position angle for the system, as shown in Figure \ref{fig:rotation2}.   We fit the following equation to the data using a non-linear least squares routine, based on Numerical Recipes MRQMIN \citep{NM}.  
\begin{equation}
\label{eq:rotation}
v(\theta) = v_{rot}\sin(\theta - \theta_0) + v_0
\end{equation}
In Equation \ref{eq:rotation}, $\theta$ is the position angle of the GCs, $v_0$ is the systemic velocity of M104 for which we adopt the robust, biweight mean velocity of the full GC sample of 1097.3 $km s^{-1}$, $v_{rot}$ is the rotation velocity of the GC system, and $\theta_0$ is the position angle of the rotation.  The best fitting curve to the full data set is shown as a solid black line in the top panel of Figure \ref{fig:rotation2}.  We determined the significance of our fit by performing a Monte Carlo simulation on a series of simulated GC systems generated by creating random pairs of the observed $\theta$ and $v$ values.  We then calculated the significance value as the percentage of the Monte Carlo cases where the resulting rotation velocity is greater than our original fit.  Thus, large values for the significance value indicate that rotation is not present within the sample.  In addition to looking at rotation over the entire GC system, we also used this method to look for evidence of radial changes in the rotation.  This was done by looking at groups of GCs in 5\arcmin~($\sim$3$R_e$ or $\sim$14 kpc) wide bins, starting from the center of the galaxy, and moving out to 20\arcmin~($\sim$12$R_e$ or $\sim$54 kpc), where the data become too sparse to perform the fit.  Best fit parameters and significance values for each of the fits are shown in Table \ref{table:rotation}.  We find no indication of significant rotation (significance values less than 1\%) in any of our bins, which suggests that as a whole, the cluster system is non-rotating.

We also looked for rotation in the red and blue subsamples within our data set.  We examined these two subpopulations for rotation using the same method described for the full sample above.  The results are also tabulated in Table \ref{table:rotation} and illustrated in the bottom two panels of Figure \ref{fig:rotation2}.  We find no significant evidence for rotation in either the red or the blue subsamples.

\subsection{Determination of the Mass Profile}
\label{section:mass profile}

\subsubsection{Velocity Dispersion Profile}
\label{section:vdisp}

\begin{figure}
\plotone{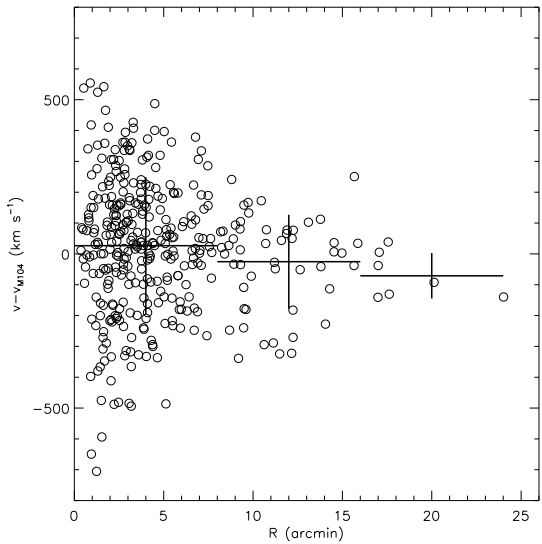}
\caption{Relative GC velocity as a function of projected radius.  GCs are indicated by the open circles.  The horizontal lines show the locations of the mean velocity for GCs in three bins divided at 8\arcmin~($\sim$4.7$R_e$ or $\sim$21.6 kpc) and 16\arcmin~($\sim$9.4$R_e$ or $\sim$43.3 kpc).  The vertical lines show the width of the average velocity dispersion for the GCs in each bin.} 
\label{fig:vdisppts}
\end{figure}

\begin{figure}
\plotone{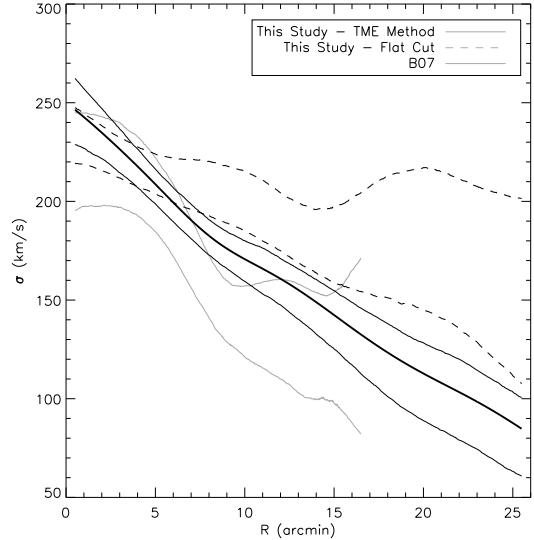}
\caption{Velocity dispersion profile, shown by the heavy solid black line, that is the result of smoothing the data in Figure \ref{fig:vdisppts} with a Gaussian with a gradually increasing width from 2\arcmin~($\sim$1$R_e$ or $\sim$5 kpc) to 4\arcmin~($\sim$2$R_e$ or $\sim$11 kpc).  The 1-$\sigma$ limits for this profile are shown by the thin solid black lines. The dashed black lines denote the velocity dispersion profile computed using the GC sample selected using the flat velocity cut.  The velocity dispersion of our final GC sample steadily decreases with radius, and does not show any signs of the flattening seen in the 1-$\sigma$ limits of the velocity dispersion profile of B07, represented by the solid gray lines.  By comparison, the velocity dispersion of the flat-cut GC sample exhibits a more flattened shape.} 
\label{fig:vdisp}
\end{figure}

Figure \ref{fig:vdisppts} shows the difference between the GC velocities and the systemic velocity of M104 plotted against the major axis distance.  In this case we have folded the data east to west in order to examine the velocity dispersion as a function of radius.  The GCs in our sample extend to a radius of 24\arcmin~($\sim$14.1$R_e$ or $\sim$64.9 kpc) from the center of the host galaxy.  We therefore divided the clusters into three radial bins at 8\arcmin~($\sim$4.7$R_e$ or $\sim$21.6 kpc), and 16\arcmin~($\sim$9.4$R_e$ or $\sim$43.3 kpc), and computed the mean velocity and average dispersion for each bin.  We find that inside 8\arcmin~the velocity dispersion is 222.1 $\pm$ 12.8 $km s^{-1}$, with a mean velocity of 1124.0 $km s^{-1}$.  In the outer two bins, the velocity dispersion decreases from 152.2 $\pm$ 21.3 $km s^{-1}$ between 8\arcmin~and 16\arcmin, to 73.7 $\pm$ 27.8 $km s^{-1}$ outside 16\arcmin.  Similarly, the mean velocity also decreases from 1072.1 $km s^{-1}$ to 1026.3 $km s^{-1}$ in the middle and outer bins, respectively.  We then computed the velocity dispersion profile by smoothing the discrete velocity profile with a Gaussian of gradually increasing width from $\sigma$ = 2\arcmin~($\sim$1$R_e$ or $\sim$5 kpc) to $\sigma$ = 4\arcmin~($\sim$2$R_e$ or $\sim$11 kpc) with increasing radius.  Figure \ref{fig:vdisp} shows the resulting smoothed profile.  The heavy solid line is the computed profile, and the thin solid lines show the 1-$\sigma$ uncertainties.  The 1-$\sigma$ dispersion profiles computed by B07 are shown as solid gray lines.  In order to compute the uncertainties in the velocity dispersion profile, we used a jackknife technique to generate 1000 new velocity dispersion profiles using 1000 subsamples of GCs each containing 270 GCs or roughly 75\% of the full sample.  These subsample profiles were then used to compute the 1-$\sigma$ range. The resulting velocity dispersion profile shows a steady decline from the center of the galaxy all the way out to the end of our data set at 25\arcmin~($\sim$15$R_e$ or $\sim$68 kpc).  Inside 10\arcmin~($\sim$6$R_e$ or $\sim$27 kpc), our velocity dispersion profile shows good agreement with the B07 profile.  However, our profile does not show the flattening seen by B07 for radii $\gtrsim$10\arcmin~($\gtrsim$6$R_e$ or $\gtrsim$27 kpc).  This is likely due to the increased number of GCs in our sample outside of 10\arcmin~($\sim$6$R_e$ or $\sim$27 kpc), which improves the sampling of the underlying dispersion profile.  We repeat the smoothing process for the GC sample selected using the flat velocity cut (illustrated by the dashed black lines in Figure \ref{fig:vdisp}).  Compared to the result for the final GC sample, this velocity dispersion profile exhibits a shape which is more flattened, although the overall trend of the velocity dispersion still decreases with radius. 

To investigate the sensitivity of the velocity dispersion profile to the location of the cut used in the TME method (see \S\ref{section:outliers}), we created additional velocity dispersion profiles for GC samples created using rejection indices of 214 and 227.  Inside of 10$\arcmin$ there is good agreement (within 1-$\sigma$) between the resulting profiles for these samples and our chosen sample.  Beyond this distance, the profiles begin to separate slightly.  The profile generated from the 227 cutoff decreases more rapidly with radius and has a 1-$\sigma$ dispersion range of 73 to 93 km s$^{-1}$ at 20$\arcmin$. Like the profile produced by the flat cut, the 214 cutoff profile has a shallower slope and has a 1-$\sigma$ dispersion range of 110 to 159 km s$^{-1}$ at 20$\arcmin$.  It should be noted that both of these profiles overlap with the velocity dispersion profile of our final GC sample shown in Figure \ref{fig:vdisppts}, and they have a similar overall shape.

 We also examined the velocity dispersion profiles of the red and blue GC subpopulations. This result is discussed in \S\ref{section:rbcomp}.

\subsubsection{Mass Profile}
\label{section:mass}

\begin{figure}
\plotone{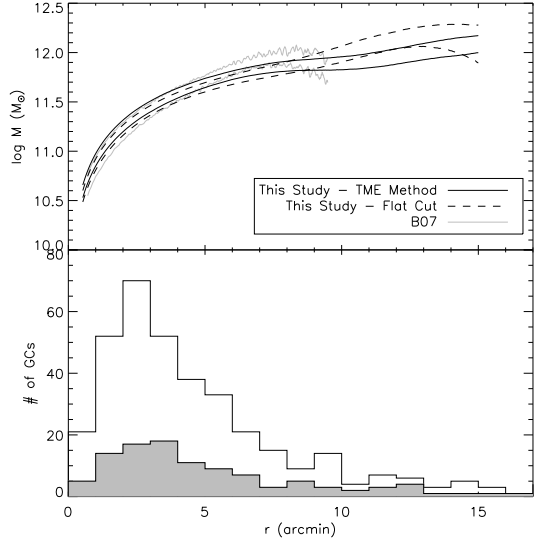}
\caption{Results of the Jeans mass modeling of the final GC sample.  The top panel shows the 1-$\sigma$ limits of our mass profile for the galaxy as solid black lines.  The dashed black lines show the result of the mass determination for GCs selected using the flat velocity cut.  The solid grey lines are the profiles determined by B07 in their study of 108 M104 GCs.  The bottom panel is a histogram showing the number of GCs with radius.  The filled histogram is the B07 sample, and the unfilled histogram is our sample.  The mass profile from our sample extends nearly 5\arcmin~($\sim$3$R_e$ or $\sim$14 kpc) farther from the galaxy center than B07.  Although there is good general agreement between our profiles and the B07 profiles, we see a slight elevation of our profiles inward of 5\arcmin~($\sim$3$R_e$ or $\sim$14 kpc).  This is most likely caused by the large increase in the number of GCs in our sample in this region.} 
\label{fig:mass}
\end{figure}

We modeled the mass profile using a spherical, isotropic Jeans mass model.  We used the smoothed velocity dispersion profiles computed above combined with the surface density profile of GCs from RZ04 as input to the model.  Figure \ref{fig:mass} shows the 1-$\sigma$ limits of our computed mass profile as solid black lines.  Similar to the smoothed velocity dispersion, the mass uncertainties were computed using 1000 jackknife samples each containing 270 GCs.  To ensure that the mass profile determined from the jackknife sample was self consistent, the velocity dispersion was first calculated from the sample and this dispersion was used to calculate the sample's mass profile.  The mass profile begins at 4.0$\times$10$^{10}$ M$_\odot$ at our innermost bin and increases to 4.9$\times$10$^{11}$ M$_\odot$ at 5\arcmin~($\sim$3$R_e$ or $\sim$14 kpc).  Beyond 5\arcmin, the mass increases more gradually, with the total mass reaching 1.3$\times$10$^{12}$ M$_\odot$ at 15\arcmin~($\sim$9$R_e$ or $\sim$41 kpc).  We do not see a flattening of the mass profile, indicating that the data do not extend to the edge of the DM halo of the galaxy.  We also show the 1-$\sigma$ limits on the mass profile computed for the flat-cut GC sample (dashed black lines).  Within the errors, the mass profiles derived from the two GC samples are consistent.  As an additional side note, the mass profiles produced from the GC samples derived using the TME method with cutoffs at 214 and 227 (see \S\ref{section:vdisp}) are also in excellent agreement with our final mass profile to within the errors, which suggests that our results are not strongly sensitive to the exact rejection index chosen.  

Also shown in Figure \ref{fig:mass} is the 1-$\sigma$ mass profile result from the B07 paper as a set of solid gray lines.  The histogram at the bottom of the figure shows the number of GCs at each radius binned in 1\arcmin~($\sim$0.5$R_e$ or $\sim$3 kpc) bins, with the shaded region showing the B07 sample for comparison.  Of significant note is that our new mass profile extends nearly $\sim$5\arcmin~($\sim$3$R_e$ or $\sim$14 kpc) farther from the galaxy center in radius than the B07 profile.  Our confidence in the mass profile beyond $\sim$15\arcmin~($\sim$9$R_e$ or $\sim$41 kpc) is low due to the low number of GC velocity measurements beyond this distance.  The effect can also be seen in the increased uncertainty in the profile.  More measurements are needed to solidify the shape of the profile for larger radii.  Overall we find good agreement between our profile and that of B07 inside 5\arcmin.  The slope of the B07 profile outside of this radius is slightly steeper than our mass profile.  At 5\arcmin, both profiles have a mass of approximately 5$\times$10$^{11}$ M$_\odot$.  However at 8\arcmin~($\sim$5$R_e$ or $\sim$22 kpc) the B07 profile is $\sim$12\% higher than our profile at the same radius.  The exact origin of the discrepancy is unclear; however it is likely related to the flattening of the B07 velocity dispersion profile at these radii.  

\subsubsection{V-band Mass-to-light Profile}
\label{section:mtol}

\begin{figure}
\plotone{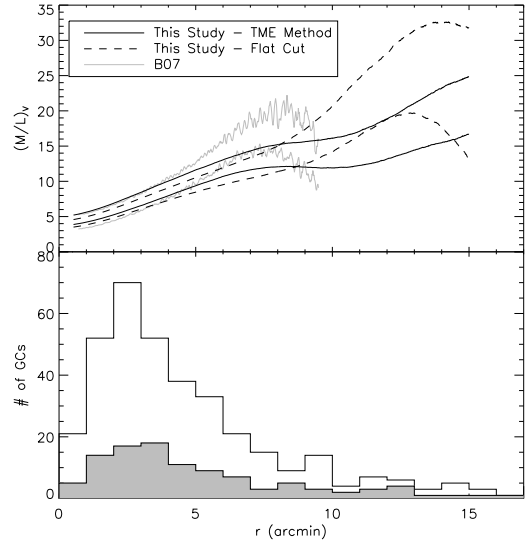}
\caption{$M/L_V$ profile for M104. The 1-$\sigma$ limits of our model are shown by the solid black lines.  The dashed black lines show the result of the mass determination for GCs selected using the flat velocity cut.  The solid grey lines are the profiles determined by B07 in their study of 108 M104 GCs.  The bottom panel is a histogram showing the number of GCs with radius.  The filled histogram is the B07 sample, and the unfilled histogram is our sample. The $M/L_V$ profile from our sample extends nearly 5\arcmin~($\sim$3$R_e$ or $\sim$14 kpc) farther from the galaxy center than B07.  Although there is good general agreement between our profiles and the B07 profiles, we see a slight elevation of our profiles inward of 5\arcmin~($\sim$3$R_e$ or $\sim$14 kpc).  This is most likely caused by the large increase in the number of GCs in our sample in this region.} 
\label{fig:mass_to_light}
\end{figure}

We next determined the $M/L_V$ profile of the galaxy by dividing our mass profile by the luminosity profile for M104 determined by \citet{Kormendy89}, which includes three components: a de Vaucouleurs profile which is the primary component, a disk component, and a nonisothermal bulge.  Each of these components is based on fits to their decomposed stellar light profiles out to roughly 5\arcmin.  The top panel of Figure \ref{fig:mass_to_light} shows the $M/L_V$ profile we computed with uncertainties shown as solid black lines.  The $M/L_V$ in the innermost bin is $\sim$4.5, and the profile rises linearly to a value of $\sim$14 at 8\arcmin~($\sim$5$R_e$ or $\sim$22 kpc).  There is a slight flattening in the profile around 10\arcmin~($\sim$6$R_e$ or $\sim$27 kpc) before it continues with a roughly linear increase to an $M/L_V$ of $\sim$19 at 15\arcmin~($\sim$9$R_e$ or $\sim$41 kpc).  It is not clear whether the flattening in the $M/L_V$ profile is real because the 1-$\sigma$ error profiles in the mass widen at this same radius.  It should be noted that the luminosity profile at these radii is an extrapolation of the \citet{Kormendy89} components.  The $M/L_V$ profile for the flat-cut GC sample, illustrated by the dashed lines, is consistent within the errors with the final GC sample.

Once again we have overplotted the result from B07 in grey.  The discrepancy between our $M/L_V$ profile and the B07 profile beyond 5\arcmin~is a direct consequence of the turnover and higher slope of the B07 mass profile at around 7\arcmin~($\sim$4$R_e$ or $\sim$19 kpc).

\section{Discussion}
\label{section:discussion}

\subsection{Comparison of the Red and Blue Sub-Populations to Other Galaxies}
\label{section:rbcomp}

Galaxies with well-studied GC kinematics are few, and are most commonly giant cluster ellipticals.  In spite of this limited sample, galaxy to galaxy comparisons of GC kinematics have begun to provide interesting results.  \citet{Hwang08} compared the kinematic properties of GCs in six well-studied giant elliptical galaxies (M60, M87, M49, NGC 1399, NGC 5128, and NGC 4636), and most recently \citet{Pota13} examined the kinematics of GCs in 12 early type galaxies (9 ellipticals and 3 S0s) as part of the SAGES Legacy Unifying Globulars and Galaxies Survey (SLUGGS).  Both of these studies found that, for their galaxies, the rotational properties of the GC systems and the GC system subpopulations were highly varied and are likely to depend on the merger history of the individual galaxy.  Numerical simulations of dissipationless mergers by \citet{Bekki05} suggest that outside a radius of $\sim$ 20 kpc ($\sim$4.3$R_e$ or $\sim$7.4\arcmin~at the distance of M104) both GC subpopulations should exhibit rotation on the order of 30 - 40 $km s^{-1}$.  However, as discussed in Section \ref{section:rotation}, we see no significant rotation in our GC sample for M104 as a whole or in the individual subpopulations.  This perhaps suggests a more complex merger history for this galaxy.

In addition to rotation, mergers can also impart differences in the overall velocity dispersion profile of the GC system.  Another prediction of the \citet{Bekki05} simulation is that the velocity dispersion profiles of the GC systems of galaxies formed by major mergers decrease as a function of radius.  In multiple-merger scenarios, they find that their modeled velocity dispersion profiles can become more flattened.  As discussed in earlier sections of this paper, the shape of the velocity dispersion profile determined from observational data is sensitive to the selection of member GCs, therefore, it is difficult to determine whether the decreasing shape of the velocity dispersion profile shown in Figure \ref{fig:vdisp} is intrinsic or a result of the GC selection process.  

\begin{figure}
\plotone{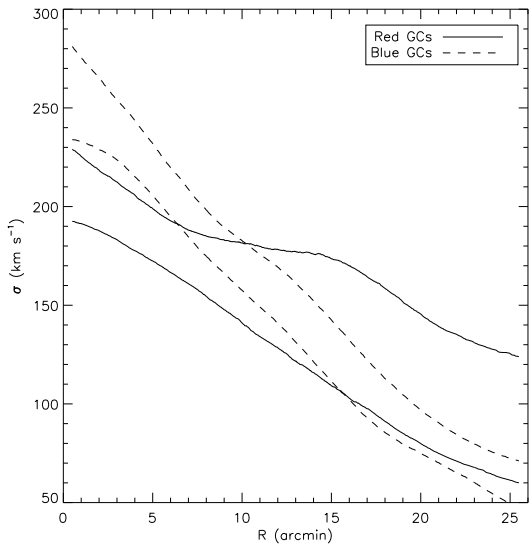}
\caption{Comparison of the velocity dispersion profiles of the red (solid lines) and blue (dashed lines) GC subpopulations.  The lines indicate the 1-$\sigma$ error profiles for each subpopulation.  Similar to \citet{Pota13} and \citet{Bekki05} We find that the central velocity dispersion of the blue GCs is higher than that of the red GCs by $\sim$ 60 $km s^{-1}$.} 
\label{fig:vdispcolor}
\end{figure}

Although the shape of the velocity dispersion profile is uncertain, we can still compare the properties of the velocity dispersion for the red and blue subpopulations.  \citet{Pota13} found observational evidence for a difference in the central velocity dispersions between the subpopulations in the GC systems of their SLUGGS galaxies.  They found that, in general, the velocity dispersion profiles for the blue GC subpopulations were higher overall than the velocity dispersion profiles of the red GCs.  Figure \ref{fig:vdispcolor} shows the smoothed velocity dispersion profiles for the red and blue GC subpopulations in our M104 sample.  Consistent with the results of \citet{Pota13} and the simulations of \citet{Bekki05} we find that the center of the velocity dispersion profile of the blue GCs is roughly 60 $km s^{-1}$ higher than that of the red GCs inside a radius of 10\arcmin($\sim$6$R_e$ or $\sim$27 kpc).   

\subsection{Comparison to Other Mass Tracers}
\label{section:other tracers}

\begin{figure}
\plotone{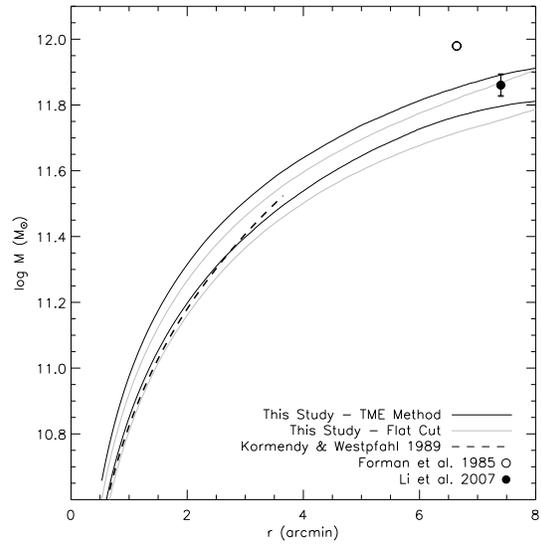}
\caption{Comparison of the inner 8\arcmin~of the GC mass profile to the mass estimates from other kinematic tracers.  The solid black and solid gray lines are the 1-$\sigma$ error profiles of our GC mass profile, with the gray lines showing the results of the GC sample selected with the flat velocity cut.  The dashed line is the mass profile from the HII study of \citet{Kormendy89}.  The circles represent the mass estimates from x-ray observations from the work of \citet{Forman85} and \citet{Li07}.  There is good overall agreement between the GC mass profile and the \citet{Kormendy89} mass profile.  The x-ray mass estimate of \citet{Li07} is in excellent agreement with the GCs.  \citet{Forman85} do not report errors on their x-ray mass estimate, but if we assume a modest uncertainty of 10\%, their result falls within the 3-$\sigma$ range of the GC mass profile.} 
\label{fig:mass_comp}
\end{figure}

The most easily observed kinematic tracers in galaxies are the stars and the gas.  Although these tracers are limited in their radial extent, it is useful to compare the results from these types of studies with the results from the GC system since they should trace the same underlying mass distribution. \citet{Kormendy89} measured the rotation curve of the stars and gas in M104 using optical spectra from the Canada-France-Hawaii Telescope and used their results to calculate the mass profile of the galaxy out to roughly 3.5\arcmin.  \citet{Bridges07} found that the \citet{Kormendy89} profile was in excellent agreement with their mass profile derived from the globular clusters (see \citealt{Bridges07} Figure 7 and associated discussion).  We also find good agreement between the \citet{Kormendy89} mass profile and our updated globular cluster mass profile.  Figure \ref{fig:mass_comp} shows the 1-$\sigma$ boundaries of our mass profile out to 8\arcmin~($\sim$4.7$R_e$ or $\sim$21.6 kpc) shown as solid black lines.  The 1-$\sigma$ mass profile boundaries for the GCs identified using a flat velocity cut are also shown with solid gray lines.  Overplotted on this figure is the mass profile of \citet{Kormendy89}, illustrated by the dashed line.  Both of our mass profiles are consistent with the \citet{Kormendy89} profile, although inward of $\sim$2\arcmin~the mass profile derived from the flat GC sample is in slightly better agreement.

X-ray emission from hot coronal gas has been predicted by galaxy formation models \citep{White78, White91}, and has been observed in many giant elliptical and S0 galaxies \citep{Forman85}.  It has also been found in a few spiral galaxies \citep{Bogdan13, Benson00}.  M104 has been shown to possess extended, diffuse x-ray emitting hot gas out to $\sim$20 kpc ($\sim$4.3$R_e$ or $\sim$7.4\arcmin) from the galaxy center \citep{Li07, Li11}.  These observations can provide additional estimates of the host galaxy mass.  Using measurements of the diffuse x-ray emission of M104 from the Einstein Observatory, \citet{Forman85} estimated a total mass for M104 of 9.5$\times$10$^{11}$ M$_\odot$ at a radius of $\sim$6.6\arcmin~($\sim$3.9$R_e$ or $\sim$17.9 kpc).  \citet{Li07} observed diffuse x-ray emission in the Sombrero using Chandra and XMM-Newton.  They measured a uniform plasma temperature of 0.6-0.7 keV extending to a radius of 20 kpc~($\sim$4$R_e$ or $\sim$7.4\arcmin) from the galaxy center.  Assuming the gas is in virial equilibrium, we calculate a total mass enclosed inside this radius between 6.7$\times$10$^{11}$ and 7.8$\times$10$^{11}$ M$_\odot$.  The masses determined from the x-ray results of \citet{Forman85} and \citet{Li07} are shown in comparison to our GC mass profile in Figure \ref{fig:mass_comp} as open and filled circles, respectively.  We plot the average value for the mass range computed from the \citet{Li07} data, with the full range indicated by the error bars.  The mass estimate from the \citet{Li07} x-ray results are in excellent agreement with our GC mass profile; however, the \citet{Forman85} mass estimate falls above our mass profile by roughly 3$\times$10$^{11}$  M$_\odot$ or approximately 5-$\sigma$.  It is difficult to judge the consistency between the \citet{Forman85} and other mass determinations due to the absence of a well-determined gas temperature.  However, we note that a modest uncertainty on the \citet{Forman85} result of 10\% would place our mass profile within 3-$\sigma$ of this result.

\section{Summary}
\label{section:results}

We have measured radial velocities for 51 previously unmeasured GCs in M104.  Combined with data from the literature, we have assembled and analyzed 360 GC velocities in this galaxy's system.  This is the largest sample of GC radial velocity measurements ever compiled for a kinematic analysis of M104.  The sample also provides double the available data beyond $\sim$10\arcmin~($\sim$6$R_e$ or $\sim$27 kpc), greatly improving the radial coverage over previous work.  As in previous studies, we find little or no evidence for rotation in the GC system as a whole, and we also do not find evidence of a significant level of rotation within the red or blue subpopulations.  We examined the velocity dispersion profile and found that the velocity dispersion decreases steadily to the edge of our available data.  We used an isotropic Jeans model to find the mass profile and $M/L_V$ profile of the GC system which extend $\sim$5\arcmin~($\sim$3$R_e$ or $\sim$14 kpc) farther from the galaxy center than previous studies.  Finally, our mass profile agrees well with masses estimated from diffuse x-ray data in the literature.

\acknowledgments
The research described in this paper was supported in part by graduate fellowships 
from the Indiana Space Grant Consortium to JLD and by an NSF Faculty Early Career Development (CAREER) award (AST-0847109) to KLR.  
We acknowledge the use of the AAOmega spectrograph at the Australian Astronomical Observatory, and we are grateful to the staff at both the AAO and at the WIYN telescope and Kitt Peak National Observatory for their assistance during our observing runs.  We would also like to offer special thanks to Sarah Brough at the AAO for her extensive help with reduction of the AAOmega data.  This research has made use of the NASA/IPAC Extragalactic Database (NED) which is operated by the Jet Propulsion
Laboratory, California Institute of Technology, under contract with the National Aeronautics and Space Administration.  Finally, we would like to thank the anonymous referee for their helpful suggestions for improving the manuscript.

\clearpage

\clearpage

\input{tab1.tex}

\input{tab2.tex}

\end{document}

%% file: tab1.tex
\begin{deluxetable}{rccccccccccc}
\tabletypesize{\scriptsize}
\tablecaption{Properties of the Target Objects with Measured Velocities}
\tablewidth{0pt}
\tablehead{
\colhead{ID} &\colhead{RA} &\colhead{Dec.} &\colhead{V} &\colhead{B$-$V} &\colhead{B$-$R} &\colhead{X} &\colhead{Y} &\colhead{R} &\colhead{$\theta$} &\colhead{$v_R$} \\
~ & (J2000) & (J2000) &~ &~ &~ & (arcmin) & (arcmin) & (arcmin) & (degrees) & ($km s^{-1}$)
} 
\startdata
RZ1003 & 12:40:41.12 & --11:30:49.99 & 19.09   & 0.92    & 1.53    & 10.21   & 6.55    & 12.13   & 32.67   & 63.0    $\pm$ 20.0   \\
RZ1005 & 12:40:41.07 & --11:54:11.87 & 19.65   & 0.79    & 1.32    & 10.19   & --16.81  & 19.66   & 301.21  & 63.0    $\pm$ 28.0   \\
RZ102 & 12:41:13.25 & --11:48:14.43 & 21.62   & 0.56    & 0.98    & 18.07   & --10.86  & 21.08   & 328.99  & 296.3   $\pm$ 11.1   \\
RZ1030 & 12:40:40.23 & --11:19:23.77 & 20.29   & 0.63    & 1.06    & 10.00   & 17.99   & 20.58   & 60.92   & 178.0   $\pm$ 47.0   \\
RZ1039 & 12:40:39.93 & --11:25:59.49 & 20.49   & 0.71    & 1.18    & 9.92    & 11.39   & 15.11   & 48.94   & 278.7   $\pm$ 6.8    \\
RZ1045 & 12:40:39.82 & --11:36:55.14 & 19.67   & 0.76    & 1.24    & 9.89    & 0.46    & 9.90    & 2.69    & 1025.5  $\pm$ 26.9   \\
RZ1059 & 12:40:39.33 & --11:41:16.8 & 19.29   & 1.07    & 1.76    & 9.77    & --3.90   & 10.52   & 338.25  & 62.0    $\pm$ 30.0   \\
RZ1060 & 12:40:39.31 & --11:52:08.0 & 19.83   & 1.07    & 1.77    & 9.76    & --14.75  & 17.69   & 303.48  & --15.0   $\pm$ 36.0   \\
RZ1061 & 12:40:39.30 & --11:45:10.86 & 20.70   & 0.65    & 1.09    & 9.76    & --7.80   & 12.49   & 321.37  & 259.3   $\pm$ 10.9   \\
RZ1070 & 12:40:39.00 & --11:55:01.95 & 19.68   & 0.73    & 1.21    & 9.68    & --17.65  & 20.13   & 298.74  & 1005.9  $\pm$ 31.6   \\
\enddata
\protect\label{table:gcdata}
\tablecomments{Table \ref{table:gcdata} is published in its entirety in the electronic edition of the AJ. A portion is shown here for guidance regarding its form and content. Photometric data provided in this table is a compilation of data from the literature, and is not available for every source.}
\end{deluxetable}

%% file: tab2.tex
\begin{deluxetable}{rcccc}
\tablecaption{Parameters for Best Fit Rotation Curves}
\tablewidth{0pt}
\tablehead{
\colhead{GC Sample} &\colhead{N} &\colhead{$v_{rot}$} &\colhead{$\theta_0$} &\colhead{Significance} \\
~ & ~ & ($km s^{-1}$) & (degrees) & (\%) 
} 
\startdata
All & 360 & 30.4 $\pm$ 22.1 & 54.7 $\pm$ 25.0 & 83.9\\
0-5\arcmin & 233 & 41.7 $\pm$ 32.8 & 43.2 $\pm$ 26.3 & 82.9 \\
5-10\arcmin & 92 & 18.6 $\pm$ 33.1 & 359.9 $\pm$ 91.3 & 24.4 \\
10-15\arcmin & 25 & 100.9 $\pm$ 41.8 & 133.8 $\pm$ 19.4 & 91.4\\
15-20\arcmin & 8 & 55.8 $\pm$ 49.2 & 198.9 $\pm$ 58.5 & 2.5 \\
Red GCs & 149 & 43.7 $\pm$ 23.9 & 342.7 $\pm$ 29.0 & 82.3\\
Blue GCs & 210 & 47.0 $\pm$ 22.1 & 80.9 $\pm$ 25.0 & 91.4 \\
\enddata
\protect\label{table:rotation}
\end{deluxetable}

%% file: M104_paper_with_referee_edits.bbl
\clearpage

\begin{thebibliography}{}

\bibitem[Alves-Brito et al.(2011)]{Deimos11}
	Alves-Brito, A., Hau, G., Forbes, D., et al. 2011, MNRAS, 417, 1823
\bibitem[Ashman \& Zepf (1998)]{Ashman98}
	Ashman, K. M., \& Zepf, S. E. 1998, Globular Cluster Systems (Cambridge: Cambridge University Press)
\bibitem[Bekki et al.(2005)]{Bekki05}
	Bekki, K., Beasley, M. A., Brodie, J. P., \& Forbes, D. A. 2005, MNRAS, 363, 1211
\bibitem[Benson et al.(2000)]{Benson00}
	Benson, A. J., Bower, R. G., Frenk, C. S., \& White, S. D. M. 2000, MNRAS, 314, 557
\bibitem[Bridges et al.(2007)]{Bridges07}
	Bridges, T., Rhode, K., Zepf, S., \& Freeman, K. 2007, AJ, 658, 980
\bibitem[Bridges et al.(1997)]{Bridges97}
	Bridges, T., Ashman, K., Zepf, S., et al. 1997, MNRAS, 284, 376
\bibitem[Brodie \& Strader(2006)]{Brodie06}
	Brodie, J. \& Strader, J. 2006, ARA\&A, 44, 193
\bibitem[Bruzual \& Charlot(2003)]{Bruzual03}
	Bruzual, G. \& Charlot, S. 2003, MNRAS, 344, 1000
\bibitem[Bogdan et al.(2013)]{Bogdan13}
	Bogdan, A., Forman, W. R., Vogelsberger, M., et al. 2013, ApJ, 772, 97
\bibitem[C\^ot\'e et al.(2001)]{Cote01}
	C\^ot\'e, P., McLaughlin, D. E., Hanes, D. A., et al. 2001, ApJ, 559, 828
\bibitem[C\^ot\'e et al.(2003)]{Cote03}
	C\^ot\'e, P., McLaughlin, D. E., Cohen, J. G., Blakeslee, J. P. 2003, ApJ, 591, 850
\bibitem[Croom et al.(2005)]{Croom05}
	Croom, S., Saunders, W., Heald, R., \& Bailey, J. 2005, The 2dfdr Data Reduction System Users Manual (Epping: AAO), http://www.aao.gov.au/AAO/2df/manual.html
\bibitem[Dirsch et al.(2003)]{Dirsch03}
	Dirsch, B., Richtler, T., Geisler, D., et al. 2003, AJ, 125, 1908
\bibitem[Dirsch et al.(2004)]{Dirsch04}
	Dirsch, B., Richtler, T., Geisler, D., et al. 2004, AJ, 127, 2114
\bibitem[Evans et al.(2003)]{Evans03}
	Evans, N. W., Perret, K. M., \& Bridges, T. J. 2003, ApJ, 583, 752
\bibitem[Forman et al.(1985)]{Forman85}
	Forman, W., Jones, C., \& Tucker, W. 1985, ApJ, 293, 102
\bibitem[Gebhardt \& Kissler-Patig(1999)]{Gebhardt99}
	Gebhardt, K. \& Kissler-Patig, M. 1999, AJ, 118, 1526
\bibitem[Hanes et al.(2001)]{Hanes01}
	Hanes, D. A., C\^ot\'e, P., Bridges, T. J., et al. 2001, ApJ, 559, 812
\bibitem[Harris (2009)]{Harris09}
	Harris, W. E. 2009, ApJ, 703, 939
\bibitem[Harris et al.(2010)]{Harris10}
	Harris, W. E., Spitler, L. R., Forbes, D. A., \& Bailin, J. 2010 MNRAS, 401, 1965
\bibitem[Hwang et al.(2008)]{Hwang08}
	Hwang, H. S., Lee, M. G., Park, H. S., et al. 2008, ApJ, 674, 869
\bibitem[Kissler-Patig et al.(1998)]{KP98}
	Kissler-Patig, M., Brodie, J. P., Schroder, L. L., et al. 1998, AJ, 115, 105
\bibitem[Kormendy \& Westpfahl(1989)]{Kormendy89}
	Kormendy, J., \& Westpfahl, D. 1989, ApJ, 338, 752
\bibitem[Kundu \& Whitmore(2001)]{Kundu01}
	Kundu, A. \& Whitmore, B. 2001, AJ, 121, 2950
\bibitem[Larsen et al.(2002)]{Larsen02}
	Larsen, S., Brodie, J., Beasley, M., \& Forbes, D. A. 2002, AJ, 124, 828
\bibitem[Larsen et al.(2001)]{Larsen01}
	Larsen, S., Forbes, D., \& Brodie, J. 2001, ApJ, 602, 705
\bibitem[Li et al.(2011)]{Li11}
	Li, Z., Jones, C., Forman, W. R., et al. 2011, ApJ, 730, 84
\bibitem[Li et al.(2007)]{Li07}
	Li, Z., Wang, Q. D., \& Hameed, S. 2007, MNRAS, 376, 960 
\bibitem[Peng et al.(2004)]{Peng04}
	Peng, E. W., Ford, H. C., Freeman, K. C. 2004, ApJ, 150, 367
\bibitem[Press et al. (1992)]{NM}
	Press, W. H., Teukolsky, S. A., Vetterling, W. T., \& Flannery, B. P. 1992, Numerical Recipes in C (2nd ed.; New York, NY: Cambridge University Press)
\bibitem[Pota et al.(2013)]{Pota13}
	Pota, V., Forbes, D. A., Romanowsky, A. J., et al. 2013, MNRAS, 428, 389
\bibitem[Rhode \& Zepf(2004)]{Rhode04}
	Rhode, K., \& Zepf, S. 2004, AJ, 127, 302
\bibitem[Schuberth et al.(2010)]{Schuberth10}
	Schuberth, Y., Richtler, T., Hilker, M., et al. 2010, A\&A, 513, 52
\bibitem[Schuberth et al.(2012)]{Schuberth12}
	Schuberth, Y., Richtler, T., Hilker, M., et al. 2012, A\&A, 544, 115
\bibitem[Spitler et al.(2006)]{Spitler06}
	Spitler, L. R., Larsen, S. S., Strader, J., et al. 2006, AJ, 132, 1593
\bibitem[Strader et al.(2011)]{Strader11}
	Strader, J., Romanowsky, A. J., Brodie, J. P., et al. 2011, ApJ, 197, 33
\bibitem[Tonry, et al.(2001)]{Tonry01}
	Tonry, J. L., Dressler, A., Blakesee, J. P., et al. 2001, ApJ, 546, 681
\bibitem[Tonry \& Davis(1979)]{TonryDavis}
	Tonry, J. \& Davis, M. 1979, AJ, 84, 1511
\bibitem[Tully et al.(2008)]{Tully08}
	Tully, B. R., Shaya, E. J., Karachentsev, I. D., et al. 2008, ApJ, 676, 184
\bibitem[van der Marel et al.(1994)]{Vandermarel94}
	van der Marel, R., Rix, H., Carter, D., et al. 1994, MNRAS, 268, 521
\bibitem[White \& Frenk(1991)]{White91}
	White, S. D. M. \& Frenk, C. S. 1991, ApJ, 379, 52
\bibitem[White \& Rees(1978)]{White78}
	White, S. D. M., \& Rees, M. J. 1978, MNRAS, 183, 341
\bibitem[Woodley et al. (2007)]{Woodley07}
	Woodley, K. A., Harris, W. E., Beasley, M. A., et al. 2007, AJ, 134, 494

\end{thebibliography}
